\documentstyle[12pt,epsf]{article}
\renewcommand{\bar}[1]{\overline{#1}}

\begin{document}

\begin{flushright} 
hep-ph/9707226\\
BIHEP-TH-97-01\\ 
\end{flushright}

\bigskip
\bigskip
\bigskip
\bigskip
\bigskip
\centerline{{\large \bf Strange Magnetic Moment 
and Isospin Symmetry Breaking}\footnote{\baselineskip=13pt
Work supported by National Natural Science Foundation of China under
Project 19445004 and Project 19605006 
and National Education Committee of China 
under Grant No.~JWSL[1995]806.}}
\vspace{62pt}

\centerline{\bf Bo-Qiang Ma}
\vspace{8pt}
\centerline{CCAST (World Laboratory), P.O.~Box 8730, Beijing 100080, China}
\centerline{Institute of High Energy Physics, Academia Sinica}
\centerline{P.~O.~Box 918(4), Beijing 100039, China\footnote{Mailing
address. E-mail: mabq@bepc3.ihep.ac.cn.}} 
\centerline{and}
\centerline{Institute of Theoretical Physics, Academia Sinica, 
Beijing 100080, China}
\vfill 
\centerline{To be published in Physics~Letters~B.} 
\vfill 
\newpage

\vfill

\vspace{40pt}
\begin{center} {\Large \bf Abstract}

\end{center}

The small mass difference $m_n-m_p=1.3$~MeV between
the proton and neutron leads to an excess of $n=\pi^- p$
over $p=\pi^+ n$ fluctuations which can be calculated
by using a light-cone
meson-baryon fluctuation model of intrinsic 
quark-antiquark pairs of the nucleon sea. The
Gottfried sum rule violation may partially be explained 
by isospin symmetry breaking
between the proton and neutron and the same effect
introduces
correction terms to the anomalous magnetic moments and the anomalous 
weak magnetic moments of the proton and neutron.
We also evaluated the strange magnetic moment of the
nucleon from the lowest strangeness $K \Lambda$ fluctuation and
found a non-trivial influence due to isospin symmetry
breaking in the experimental measurements of the strange magnetic
moment of the nucleon.

\vfill
\centerline{PACS numbers: 13.40.Em, 11.30.Hv, 12.39.Ki, 14.20.Dh} 
\vfill
\newpage

\section{Introduction}

There have been considerable theoretical investigations and
experimental activities on the strange content of the nucleon
in recent years. The strange content of the nucleon
arises from the non-valence sea quarks and provides
a direct window into the nonperturbative QCD nature of the
quark sea in the quantum bound-state structure of the hadronic
wavefunctions \cite{Bro81,Sig87,Bur92,Bro96}. 
There have been several novel and unexpected 
features or discoveries related to the strange content of the nucleon,
such as the significant strangeness content 
stemming from the
pion-nucleon sigma term \cite{Sigma}, the anti-polarized strange quark sea
in the proton from the Ellis-Jaffe sum rule violation \cite{Sea}, 
and the strange quark-antiquark asymmetry in the nucleon
sea from two different determinations of the strange quark
distributions in the nucleon \cite{Bro96,Bro97}. Recently, the strange
magnetic moment of the nucleon has also received extensive attention
both theoretically and experimentally. 

The strange magnetic moment is closely related to
the sea quark-antiquark asymmetry of the nucleon:
A non-zero strange magnetic moment
is a direct reflection of the strange-antistrange asymmetry
in the nucleon sea \cite{Bro97,Ji95}. 
Experimental measurements of the strange magnetic
moment by means of parity-violating
electron scattering have been suggested \cite{SMM,exp} 
and the first measurement by the
SAMPLE Collaboration \cite{Sample} found
\begin{eqnarray}
G^s_M(Q^2=0.1{\rm GeV}^2)=+0.23\pm 0.37 \pm 0.15 \pm 0.19 ~{\rm n.m.} .
\nonumber
\end{eqnarray}       
In many earlier 
theoretical discussions, the strange magnetic moment was
predicted to have a large negative value, for example,
of the order $-0.3~{\rm n.m.}$ \cite{For94}. However,
there have been recent discussions about the possibility
of a small negative or even positive 
strange magnetic moment \cite{small}, influenced by the progress 
of experimental measurements.

The purpose of this paper is to investigate possible 
uncertainties in the experimental measurements 
of the strange magnetic moment. 
We will show that there are terms
due to isospin symmetry breaking
in the anomalous magnetic moments 
and the anomalous weak magnetic moments
of the proton and neutron obtained by taking
into account the small mass difference $m_n-m_p=1.3$ MeV
in the non-perturbative meson-baryon fluctuations
from the intrinsic quark-antiquark pairs of the nucleon sea.
Such terms are related to the Gottfried sum rule violation 
which may partially be explained by
isospin symmetry breaking between the proton sea and
the neutron sea \cite{Ma92,Ma93} and the size of these
terms will be estimated by using
a light-cone model of nonperturbative meson-baryon 
fluctuations \cite{Bro96}. It will be shown that these correction
terms have a non-trivial influence on the 
extraction of the strange magnetic moment of the nucleon from
experimental measurements.  

\section{The strange magnetic moment measurements}

The analyses of experimental measurements of 
the strange magnetic moment are
based on the following equations:
the anomalous magnetic moments of the proton and neutron
\begin{equation} 
G^p_M=\frac{2}{3} G^u_M-\frac{1}{3} G^d_M-\frac{1}{3} G^s_M,
\label{GPM0}
\end{equation} 
\begin{equation}
G^n_M=\frac{2}{3} G^d_M-\frac{1}{3} G^u_M-\frac{1}{3} G^s_M,
\label{GNM0}
\end{equation}
and the anomalous weak magnetic moments 
(via the neutral weak vector current matrix elements) of the proton
and neutron
\begin{equation}
G^{Zp}_M= e_u^Z G^u_M +e_d^Z G^d_M+ e_s^Z G^s_M,
\label{GZPM0}
\end{equation}
\begin{equation}
G^{Zn}_M= e_u^Z G^d_M +e_d^Z G^u_M+ e_s^Z G^s_M,
\label{GZNM0}
\end{equation}
where
$e_u^Z=(\frac{1}{4} - \frac{2}{3} \sin ^2\theta_W)$ 
and $e_d^Z=e_s^Z=(-\frac{1}{4} +\frac{1}{3}
\sin ^2\theta_W)$.

Combining Eqs.(\ref{GPM0}), (\ref{GNM0}) and (\ref{GZPM0}),
one has
\begin{equation}
G^{Zp}_M=(\frac{1}{4}-\sin ^2 \theta_W)G^p_M-
\frac{1}{4}G^n_M-\frac{1}{4}G^s_M,
\label{MGSM0}
\end{equation}
from which one can extract the strange magnetic moment $G^s_M$
from the measurable $G^{Zp}_M$ and the known experimental data
of $G^p_M$ and $G^n_M$. 

One important and strong assumption in the above equations
is the isospin symmetry between the proton and neutron.
Thus a source of isospin symmetry breaking between
the proton and neutron may introduce uncertainties in the extracted
value
of the strange magnetic moment $G^s_M$ based on the above equations.

\section{Isospin symmetry breaking}

The discovery of the Gottfried sum rule (GSR) \cite{GSR} 
violation by the New Muon Collaboration (NMC) \cite{NMC91} motivated 
studies on the flavor content of the nucleon sea
\cite{Bro96,Ma92,Pre91,Pi}.
It is commonly taken for granted that this violation
is due to the flavor asymmetry between the $u\bar u$ and $d \bar d$
quark pairs in the nucleon sea while still preserving
the isospin symmetry between the proton and the neutron
\cite{Pre91,Pi}.
Nevertheless, it has been suggested \cite{Ma92} that the GSR violation
could alternatively be explained by the isospin symmetry
breaking between the proton and the neutron while
preserving the flavor distribution symmetry in the nucleon sea. 
The flavor asymmetry between the $u$ and $d$ sea quarks
is likely the main source by the excess of the intrinsic
$d\bar d$ pairs over $u \bar u$ pairs in the proton sea through
$p(uud)=\pi^+(u\bar d) n(udd)$ over 
$p(uud)=\pi^- (d\bar u)\Delta^{++}(uuu)$ meson-baryon fluctuations
\cite{Bro96,Pi}. 
But one can not exclude the
possibility that the isospin symmetry breaking partially
contributes to the GSR violation and suggestions have been
made \cite{Ma92,Ma93} as to how to distinguish between the two possible 
explanations.

We will show in the following that the same mechanism leading
to the excess of $d \bar d$ pairs over $u \bar u$ pairs in the proton
sea could also lead to a small excess of the lowest meson-baryon
fluctuation $n(udd)=\pi^-(d\bar u) p(uud)$ for the neutron
over $p(uud)=\pi^+(u\bar d) n(udd)$ for the proton within
a light-cone model of energetically-favored meson-baryon
fluctuations \cite{Bro96}. In a strict sense, we still
lack a basic theory to produce the probabilities of 
fluctuations due the the non-perturbative nature of those
fluctuations. However, the probabilities for such fluctuations
can be inferred from experimental measurements of physical
quantities related to those fluctuations. For example,
the amplitude of 
the lowest fluctuation state $p=\pi^+ n$ for the proton is
of the order 15\% as estimated from the measured Gottfried sum \cite{Bro96}
and the amplitude of lowest strangeness fluctuation state $p=K^+ \Lambda$
is of the order 5\% from re-producing empirical measurements 
related to the strange content of the nucleon \cite{Bro96,Song}.

From the uncertainty principle, we can also estimate 
the relative probabilities of two meson-baryon states by 
comparing their off-shell light-cone energies 
with the static nucleon bound state. 
For example, we can use the light-cone Gaussian type wavefunction
\cite{Bro96,Bro97,BHL}
\begin{equation} 
\psi_{{\rm Gaussian}}({\cal M}^2)=A_{{\rm Gaussian}}
\  {\rm exp} [-({\cal M}^2-m_N^2)/8\alpha^2],
\end{equation} 
where 
${\cal M}^2=\sum_{i=1}^{2} \frac{{\bf k}^2_{\perp i}+m_i^2}{x_i}$ 
is the invariant mass for the meson-baryon state, $m_N$ is the
physical mass of the nucleon, and $\alpha$ sets the characteristic 
internal momentum scale, with the same normalization constant
$A_{{\rm Gaussian}}$ to evaluate the relative probabilities
of two meson-baryon fluctuation states.
With the parameter value $\alpha=330$ MeV as previously 
adopted \cite{Bro96}, 
and with the physical masses $m_p=938.27$, $m_n=939.57$, and
$m_{\pi}=139.57$ MeV for the proton, neutron, and charged pion \cite{PDG}, 
we find the ratio of the fluctuation probabilities 
\begin{equation}
r^{\pi}_{p/n} = P(p=\pi^+ n)/P(n=\pi^- p)=0.986, 
\end{equation}
which is equivalent to an excess of 0.2\% of $n=\pi^- p$
over $p=\pi^+ n$ fluctuations assuming
$P(p=\pi^+n) \approx P(n=\pi^- p) \approx 0.15$.
There are still uncertainties in the evaluation of the
ratio $r^{\pi}_{p/n}$. For example, the parameter $\alpha$
reflects the relative internal motions of the pions
around the baryon and might be smaller, e.g.
$\alpha \approx 200$ MeV, compared to $\alpha=330$ MeV due to the
small pion mass $m_{\pi}$,
and the Coulomb attraction between $\pi^-$ and $p$
in the $n=\pi^- p$ state may cause slightly larger
relative motions of pions (e.g. $\alpha=205$ MeV) 
than that in the $p=\pi^+ n$
state (e.g. $\alpha=200$ MeV) from the uncertainty principle. 
With the parameters
$\alpha=205$ MeV for the neutron and $\alpha=200$ MeV
for the proton, we find the ratio $r^{\pi}_{p/n}=0.820$, which
is equivalent to an excess of 3\% of $n=\pi^- p$
over $p=\pi^+ n$ fluctuations. 
In a strict sense, there is no reason to assume a
single value for the normalization constant $A_{\rm Gaussian}$
and this may introduce further
uncertainties about the amplitude of isospin symmetry breaking. 
Therefore the excess
of $n=\pi^- p$
over $p=\pi^+ n$ fluctuation probabilities lies in the range
\begin{equation}
\delta P^{\pi}=P(n=\pi^- p)-P(p=\pi^+n)=0.002 \to 0.03 
\label{enp}
\end{equation}
from a crude model estimation.

There are also neutral fluctuations (i.e., the fluctuation
of chargeless mesons $\pi^0$ {\it et al.}) in the nucleons but 
the isospin symmetry breaking arising from those
neutral fluctuations can be shown to be negligibly small
by using similar estimations to those above. 
The ratio of the strangeness fluctuations
\begin{equation}
%r^K_{p/n}=P(p=K^+\Lambda)/P(n=K^- \Lambda)=0.994, 
 r^K_{p/n}=P(p=K^+\Lambda)/P(n=K^0 \Lambda)=1.02, 
\end{equation}
which is equivalent to an excess of 
%0.03\% of $n=K^- \Lambda$ over $p=K^+ \Lambda$ 
  0.1\% of $p=K^+ \Lambda$ over $n=K^0 \Lambda$  
fluctuations assuming
%$P(p=K^+\Lambda) \approx P(n=K^- \Lambda) \approx 0.05$.
 $P(p=K^+\Lambda) \approx P(n=K^0 \Lambda) \approx 0.05$.
Thus we can neglect the isospin symmetry breaking from the strangeness
fluctuations.
We also ignore here the possible isospin symmetry breaking 
in the valence quarks between the proton and neutron \cite{CSB}.
Thus the excess of 
$n=\pi^- p$ over $p=\pi^+ n$ fluctuation states
due to the small mass difference $m_n-m_p=1.3$ MeV
seems to be an important source for the isospin symmetry
breaking between the proton and the neutron. If the GSR violation 
is entirely due to isospin symmetry breaking
between the proton and the neutron, then 
the NMC measurement $ S_G=\frac{1}{3}+\frac{10}{9}
\int_0^1{\rm d} x [O_q^p(x)-O_q^n(x)] =0.235 \pm 0.026 $ \cite{NMC91}
implies 
\begin{equation}
\int_0^1{\rm d} x [O_q^n(x)-O_q^p(x)]=\frac{3}{10}-\frac{9}{10} S_G=
0.088 \pm 0.023,
\end{equation}
which can be considered as the excess of quark-antiquark 
pairs in the neutron over those in the proton \cite{Ma92,Ma93}.
Eq.~(\ref{enp}) means that the isospin symmetry breaking
between the proton and the neutron
contributes only a small part of the GSR violation 
or the sea of the pion also contributes some part to the GSR violation.

\section{The measured ``strange magnetic moment"}

If we take into account the correction due to isospin
symmetry breaking, the anomalous magnetic moment of the neutron should
be written as
\begin{equation}
G^n_M=\frac{2}{3} G^d_M-\frac{1}{3} G^u_M-\frac{1}{3} G^s_M
+\delta G^n_M.
\label{GNM}
\end{equation}
Combining Eqs.(\ref{GPM0}), (\ref{GZPM0}) and (\ref{GNM}),
one has
\begin{eqnarray}
G^{Zp}_M
&=&(\frac{1}{4}-\sin ^2 \theta_W)G^p_M-
\frac{1}{4}(G^n_M-\delta G^n_M)-\frac{1}{4}G^s_M 
\nonumber\\
&=&(\frac{1}{4}-\sin ^2 \theta_W)G^p_M-
\frac{1}{4}G^n_M-\frac{1}{4}\hat{G}^s_M,
\label{MGSM}
\end{eqnarray}
from which we know that the ``strange magnetic moment" measured
from Eq.~(\ref{MGSM0}) is actually
\begin{equation}
\hat{G}^s_M=G^s_M-\delta G^n_M.
\label{GSM1}
\end{equation}

We now roughly estimate the correction term due to isospin
symmetry breaking in the neutron anomalous magnetic moment. 
In the light-cone meson-baryon fluctuation model \cite{Bro96}, 
the total angular momentum space wavefunction of the energetically
most favored
intermediate 
$N=MB$ state in the center-of-mass reference frame should be 
\begin{eqnarray}
\left|J=\frac{1}{2},J_z=\frac{1}{2}\right\rangle
&=&\sqrt{\frac{2}{3}} \left|L=1,L_z=1\right\rangle\
\left|S^B=\frac{1}{2},S^B_z=-\frac{1}{2}\right\rangle\nonumber\\
&-&\sqrt{\frac{1}{3}}\
\left|L=1,L_z=0\right\rangle \
\left|S^B=\frac{1}{2},S^B_z=\frac{1}{2}\right\rangle\ .
\end{eqnarray}
After taking into account the contributions
from the baryon spin and the orbital motions of the meson and
baryon, we have the magnetic moment of the meson-baryon 
state 
\begin{equation}
\mu^N_{MB}=\left\langle 2 S^B_z \right\rangle \mu_{B}  
+e_B \left\langle L^B_z \right\rangle \mu^B_L
+ e_M \left\langle L^M_z \right\rangle \mu^M_{L},
\label{gmb}
\end{equation}
where $\mu_{B}$ is the baryon magnetic moment,
$\left\langle 2 S^B_z \right\rangle=-\frac{1}{3}$ is the fractional
spin contribution to the nucleon from the baryon,
$e_B$ and $e_M$ are the electric charges of the baryon and meson,
$\left\langle L^B_z \right\rangle=\frac{m_M}{m_M+m_B}
\left\langle L_z \right\rangle$ 
and $\left\langle L^M_z \right\rangle=\frac{m_B}{m_M+m_B}
\left\langle L_z \right\rangle$ are the orbital angular momenta
of the baryon and meson with % relative orbital angular moment  
$\left\langle L_z \right\rangle=\frac{2}{3}$,
$\mu^B_L=\frac{m_N}{m_B} \mu_N$ and $\mu^M_L=\frac{m_N}{m_M} \mu_N$
are the contributions from the orbital motions of the baryon
and meson with unit charge and unit orbital angular momentum.
For the $n=\pi^- p$ state, we have
\begin{equation}
\mu^{n}_{\pi^- p}=-\frac{1}{3}\mu_p+\left\langle L^p_z \right\rangle
\mu^p_L-\left\langle L^{\pi}_z \right\rangle \mu^{\pi}_{L}.
\label{mul} 
\end{equation}
Thus the correction term due to isospin symmetry breaking 
to the neutron anomalous magnetic moment is 
\begin{eqnarray}
\delta G^n_M
&=& \delta P^{\pi}(\mu^n_{\pi^- p}-\mu'_{n})
\label{dgn}
\end{eqnarray}
where $\mu'_{n}=\mu_n-\delta G^n_M$ is the neutron magnetic
moment without isospin symmetry breaking, but as an approximation,
we may first use 
the physical $\mu_{n}$ instead.
By using the experimental data $\mu_{p}=2.793$ and
$\mu_{n}=-1.913~{\rm n.m.}$
\cite{PDG},   
we have    
\begin{equation}
\delta G^n_M  \approx - 0.006 \to
-0.088~{\rm n.m.}
\label{dgnf}
\end{equation}
corresponding to 
$\delta P^{\pi} \approx 0.2 \to 3 \%$ 

As we have shown above, the strangeness fluctuations
do not introduce isospin symmetry breaking correction
terms to the magnetic moments for the proton and neutron.
We also consider only the lowest non-neutral strangeness
fluctuation $K \Lambda$ state, since higher fluctuations
might be suppressed due to larger 
off-shellness \cite{Bro96}.
In the constituent quark model, the spin of $\Lambda$ is provided
by its strange quark and the expectation value of the
spin of the antistrange quark in
the $K$ is zero.
This introduces a quark-antiquark asymmetry in the spin
and momentum distributions between the strange and antistrange
quarks \cite{Sig87,Bur92,Bro96}.  
Since the strange quark is responsible for the 
magnetic moment of the $\Lambda$, 
the strangeness contribution to the magnetic moment 
comes from the strange quark in the $\Lambda$ component 
and the orbital motions of the strange quark in the $\Lambda$
and anti-strange quark in the $K$ components: 
\begin{equation}  
\mu_{s}=
[\left\langle 2 S^{\Lambda}_z \right\rangle \mu_{\Lambda}  
+e_s \left\langle L^{\Lambda}_z \right\rangle \mu^{\Lambda}_L
+ e_{\bar s} \left\langle L^{K}_z \right\rangle \mu^K_{L}]P_{\Lambda
K},
\end{equation}
where $\mu_{\Lambda}=-0.613\pm 0.004 ~{\rm n.m.}$ 
is the $\Lambda$ magnetic moment
and $P_{\Lambda K}$ is the probability of finding the
$\Lambda K$ state in the nucleon and is of the order of 5\%. 
Thus the strange magnetic moment $G^s_M$ is 
\begin{equation}
G^s_M=\mu_{s}/(-1/3)
=[\mu_{\Lambda}+\left\langle L^{\Lambda}_z \right\rangle \mu^{\Lambda}_L
-\left\langle L^{K}_z \right\rangle \mu^K_{L}]P_{\Lambda K}
\approx -0.066 ~{\rm n.m.},
\end{equation}
which is with much smaller magnitude than most earlier 
theoretical predictions
\cite{For94}. 
We also note that considerations of higher strangeness
fluctuations in some models could 
reduce the strange magnetic moment to 
smaller negative or even positive values \cite{small}.
The isospin symmetry breaking term $\delta G^s_M$ in 
Eq.~(\ref{GSM1}) causes an increase of the order 
\begin{equation}
-\delta G^s_M \approx 0.006 \to 0.088~{\rm n.m.}
\end{equation}
to the actual strange magnetic
moment $G^s_M$ and has a non-trivial influence on the
experimental measurements of $G^s_M$.
Therefore the measured $\hat{G}^s_M$  could range from a  
small negative to even a positive value within our simple estimation. 

We turn our attention to the isospin symmetry breaking terms
in the anomalous weak magnetic moment $G^{Zn}_M$.
If there are no isospin symmetry breaking terms, 
combining Eqs.(\ref{GPM0}), (\ref{GNM0}) and
(\ref{GZNM0})
we have
\begin{equation}
G^{Zn}_M=(\frac{1}{4}-\sin ^2 \theta_W)G^n_M-
\frac{1}{4}G^p_M-\frac{1}{4}G^s_M,
\label{MGSM0N}
\end{equation}
which is similar to Eq.~(\ref{MGSM0}). We can use this to 
extract the strange magnetic moment $G^s_M$
from the measurable $G^{Zn}_M$ and the known experimental data
of $G^p_M$ and $G^n_M$. 
By taking into account the corrections due to isospin symmetry
breaking, Eq.~(\ref{MGSM0N}) should be
changed to
\begin{eqnarray}
G^{Zn}_M
&=&(\frac{1}{4}-\sin ^2 \theta_W)(G^n_M-\delta G^n_M)
-\frac{1}{4}G^p_M-\frac{1}{4}G^s_M + \delta G^{Zn}_M
\nonumber\\
&=&(\frac{1}{4}-\sin ^2 \theta_W)G^n_M
-\frac{1}{4}G^p_M-\frac{1}{4}\tilde{G}^s_M, 
\label{MGSMN}
\end{eqnarray}
where the ``strange magnetic moment" measured
from Eq.~(\ref{MGSM0N}) is actually
\begin{equation}
\tilde{G}^s_M=G^s_M+(1-4\sin ^2 \theta_W)\delta G^n_M-4 \delta
G^{Zn}_M.
\label{GSM2}
\end{equation}

Similarly to Eq.~(\ref{gmb}), we can write the 
weak magnetic moment of the meson-baryon state
\begin{equation}
\mu^Z_{MB}=
\left\langle 2 S^B_z \right\rangle \mu^Z_{B}  
+e^Z_B \left\langle L^B_z \right\rangle \mu^B_L
+ e^Z_M \left\langle L^M_z \right\rangle \mu^M_{L},
\label{gzmb}
\end{equation}
where $e^Z_B=2 e^Z_u+e^Z_d$ for $p$ and 
$e^Z_M=e^Z_d - e^Z_u$ for $\pi^-(d \bar u)$.
Similarly to Eq.~(\ref{dgn}),
the correction term 
to the anomalous
weak magnetic moment of the neutron 
due to isospin symmetry breaking is 
\begin{eqnarray}
\delta G^{Zn}_M
&=& (\mu^{Zn}_{\pi^- p}-\mu'^{Z}_n)\delta P^{\pi}
\nonumber\\
&=&\delta P^{\pi}
[-\frac{1}{3}\mu^{Z}_p+
(\frac{1}{4}-\sin ^2 \theta_W)
\left\langle L^{p}_z \right\rangle \mu^{p}_{L}
-(\frac{1}{2} -\sin ^2 \theta_W )
\left\langle L^{\pi}_z \right\rangle \mu^{\pi}_{L}-\mu'^{Z
}_n],
\label{dgzn}
\end{eqnarray}
where $\mu'^{Z}_n$ would be $\mu^{Z}_n$ without
isospin symmetry breaking  
and $\sin^2 \theta_w=0.2315$ \cite{PDG}.
After some calculation we find
\begin{equation}
\delta G^{Zn}_M 
\approx -0.001 \to -0.015~{\rm n.m.}
\end{equation} 
corresponding to 
$\delta P^{\pi} \approx 0.2 \to 3 \%$. Therefore
the correction term in Eq.~(\ref{GSM2}) causes an increase
of the measured ``strange magnetic moment"  
$\tilde{G}^s_M$ in a range
\begin{equation}
(1-4\sin ^2 \theta_W)\delta G^n_M-4 \delta
G^{Zn}_M \approx 0.004 \to 0.054~{\rm n.m.}
\end{equation}
compared to the actual strange magnetic moment $G^s_M$. 
 
In a strict sense, it is difficult to have
a model-independent measurement of the strange magnetic
moment $G^s_M$ without isospin symmetry breaking corrections
from the anomalous magnetic moments and 
the anomalous weak magnetic moments
of the proton and neutron if other
measurements of physical quantities related to the
hadronic vector matrix elements are not involved. 
However, with the help
of model-dependent theoretical connection 
between $\delta G^n_M$ and $\delta G^{Zn}_M$
we may extract the strange magnetic moment $G^s_M$
from the measured $\hat{G}^s_M$ and $\tilde{G}^s_M$.
For example, combining Eqs.~(\ref{GPM0}), (\ref{GZPM0}),
(\ref{GSM1}), and (\ref{GSM2}), supplied  
with Eqs.~(\ref{dgn}) and (\ref{dgzn}), one could 
constrain the uncertainties caused by $\delta G^n_M$ and $\delta G^{Zn}_M$
and determine
the quark matrix elements $G^u_M$, $G^d_M$, and $G^s_M$.
Other measurements of the strange vector 
form factors have also been suggested \cite{Alb96}.
Thus the experimental measurements
of the neutral weak vector matrix elements of nucleons,
both undertaken and proposed, 
will play an important role in deepening our understanding 
concerning the strange content of the nucleon.

\section{Conclusion}

We have shown that there are terms
due to isospin symmetry breaking
in the anomalous magnetic moments and the anomalous weak magnetic moments
of the proton and neutron as estimated their size
by using a
light-cone meson-baryon fluctuation model of the 
intrinsic sea quark-antiquark pairs.
Such terms are related to the Gottfried sum rule violation 
of
isospin symmetry breaking between the proton sea and
the neutron sea and the same effect might cause a non-negligible correction 
to the actual strange magnetic moment derived from the measurements.
Further theoretical and experimental work is needed
to constrain the uncertainties in the experimental
measurements of the strange magnetic moment of the nucleon.

\vspace{12pt}

\noindent
{\bf Acknowledgements}

I would like to thank V.~Barone, H.~Fokel, 
%and W.~Melnitchouk 
W.~Melnitchouk, and Mu-Lin Yan 
for helpful conversations.
I am greatly indebted to S.J.~Brodsky for his simulating
and profitable discussions and suggestions during this work.

\end{document}